\documentstyle[aps,preprint]{revtex}

\begin{document}

\preprint{EFUAZ FT-99-78}

\title{Comment on the ``Maxwell Equations as the One-Photon
Quantum Equation" by A. Gersten [Found. Phys. Lett. {\bf 12},
pp. 291-298 (1999)]
\thanks{Submitted to ``Foundation of Physics Letters"}}

\author{{\bf Valeri V. Dvoeglazov}}

\address{Escuela de F\'{\i}sica, Universidad Aut\'onoma de Zacatecas\\
Apartado Postal C-580, Zacatecas 98068 Zac., M\'exico\\
E-mail: valeri@ahobon.reduaz.mx\\
URL: http://ahobon.reduaz.mx/\~~valeri/valeri.htm
}

\date{November 13, 1999}

\maketitle

\medskip

\begin{abstract}
We show that the Gersten derivation of Maxwell equations
can be generalized. It actually leads to additional solutions
of `$S=1$ equations'. They follow directly from previous
considerations by Majorana, Oppenheimer, Weinberg and Ogievetskii and
Polubarinov.  Therefore, {\it generalized} Maxwell equations should be
used as a guideline for proper interpretations of quantum theories.
\end{abstract}

\bigskip
\bigskip

In the paper~\cite{Gers} the author studied the matrix representation
of the Maxwell equations, both the Faraday and Ampere laws and the Gauss
law.  His consideration is based on the equation (9)
\begin{equation}
\left ( {E^2\over c^2}   - \vec{\bf p}^{\,2} \right ) \vec {\bbox \Psi}
= \left ( {E\over c} I^{(3)} -\vec{\bf p}\cdot \vec{\bf S} \right )
\left ( {E\over c} I^{(3)} +\vec{\bf p}\cdot \vec{\bf S} \right )
\vec {\bbox \Psi} - \pmatrix{p_x\cr p_y \cr p_z\cr}
\left (\vec {\bf p}\cdot \vec {\bbox \Psi}\right ) =0
\quad \mbox{Eq. (9) of ref. [1]}
\nonumber
\end{equation}
Furthermore, he claimed that the solutions to this equation should be
found from the set
\begin{eqnarray}
&&\left ( {E\over c} I^{(3)} +\vec {\bf p}\cdot \vec {\bf S} \right )
\vec{\bbox\Psi} =0\quad\qquad\mbox{Eq. (10) of ref. [1]}\nonumber\\
&&\left (\vec {\bf p}\cdot \vec {\bbox \Psi}\right ) =0
\qquad\qquad\mbox{Eq. (11) of ref. [1]}\nonumber
\end{eqnarray}
Thus, Gersten concluded that the equation (9) is equivalent
to the Maxwell equations (10,11).
As he also correctly indicated, such a formalism for describing
$S=1$ fields  has been considered by several authors before. See, for
instance,~\cite{Opp,Good,Wein,Gian,Fushchich,Ahl,Dvo,DvoN,Bruce};
those authors mainly considered the  dynamical Maxwell equations
in the matrix form.

However, we straightforwardly note
that  the equation (9) of~\cite{Gers} is satisfied also under
the choice\footnote{We leave the analysis of possible
functional non-linear (in general) dependence of $\chi$ and $\partial_\mu
\chi$ on the higher-rank tensor fields for future publications.}
\begin{eqnarray}
&&\left ( {E\over c} I^{(3)} + \vec{\bf p}\cdot \vec {\bf S} \right )
\vec{\bbox\Psi} =\vec{\bf p} \chi\\
&&\left (\vec {\bf p}\cdot \vec {\bbox \Psi}\right ) ={E\over c} \chi\, ,
\end{eqnarray}
with some arbitrary scalar function $\chi$ at this stage.
This is due to the fact that\footnote{See the explicit form
of the angular momentum matrices in Eq. (6) of the Gersten paper.}
$$ (\vec {\bf p}\cdot \vec{\bf S})^{jk} \vec {\bf p}^k
= i\epsilon^{jik} p^i p^k \equiv 0 $$
(or after quantum operator substitutions \, $\mbox{rot} \,\mbox{grad}
\chi =0$).  Thus, the generalized coordinate-space  Maxwell equations
follow after the similar procedure as in~[1]:
\begin{equation}
{\bbox \nabla}\times {\bf
\vec{E}}=-\frac{1}{c}\frac{\partial {\bf \vec{B}}}{\partial t} + {\bbox
\nabla} {\it Im} \chi  \label{1}
\end{equation}

\begin{equation}
{\bbox \nabla }\times {\bf \vec{B}}=\frac{1}{c}\frac{\partial {\bf
\vec{E}}}{\partial t}  +{\bbox \nabla} {\it Re} \chi\label{2}
\end{equation}

\begin{equation}
{\bbox \nabla }\cdot {\bf \vec{E}}=-{1\over c} {\partial \over \partial t}
{\it Re}\chi \label{3}
\end{equation}

\begin{equation}
{\bbox \nabla }\cdot {\bf \vec{B}}=
{1\over c} {\partial \over \partial t} {\it Im} \chi .  \label{4}
\end{equation}
If one assumes that there are no monopoles, one may suggest that $\chi (x)$
is a {\it real} field and its derivatives play the role of charge and
current densities.  Thus, surprisingly, on using the Dirac-like
procedure\footnote{That is to say, on the basis of the relativistic
dispersion relations $$ \left( E^{2}-c^{2}{\bf
\vec{p}}^{2}-m^{2}c^{4}\right) \Psi =0,\qquad \mbox{(1) of
ref.~\cite{Gers}}.$$} of derivation of ``free-space" relativistic quantum
field equations, Gersten might in fact have come to the {\it
inhomogeneous} Maxwell equations!\footnote{One can also substitute
$-(4\pi i\hbar /c) \vec{\bf j}$ and $(-4\pi i\hbar) \rho$ in
the right hand side of (2,3) of the present paper  and obtain equations
for the current and the charge density
\begin{eqnarray} && {1 \over
c} \vec{\bbox\nabla}\times \vec{\bf j} =0\,,\\ && {1\over c^2} {\partial
\vec {\bf j} \over \partial t} +\vec{\bbox\nabla} \rho =0\,,
\end{eqnarray}
which coincide with equations (13,17) of~[9b]. The
interesting question is: whether such defined $\vec{\bf j}$ and $\rho$ may
be related to $\partial_\mu \chi$.} Furthermore, I am not aware of any
proofs that the scalar field $\chi (x)$ should be firmly connected with
the charge and current densities, so there is sufficient room for
interpretation. For instance, its time derivative and gradient may also be
interpreted as leading to the 4-vector potential. In this case, we need
some {\it mass}/{\it length} parameter as in~[11a,d]. Both these
interpretations were present in the literature~\cite{DvoN,add}
(cf. also~\cite{lh}).

Below we discuss only one aspect of the above-mentioned
problem with additional scalar  field and its derivatives
in generalizations  of the Maxwell formalism. It is connected with the
concept of {\it notoph} of Ogievetski\u{\i} and Polubarinov
(in the US journal literature it is known as the Kalb-Ramond
field).\footnote{In my opinion, Prof. S. Weinberg~\cite[p.208]{WB}
confirmed this idea when considering a spin-0 4-vector field
in his famous book.} The
related problem of misunderstandings of the Weinberg theorem
$B-A=\lambda$ is slightly discussed too; $A$ and $B$ are eigenvalues of
angular momenta corresponding to certain representation of the Lorentz
Group, $\lambda$ is the helicity~\cite[p.B885]{Wein}.

Actually, after perfoming the Bargmann-Wigner procedure of description of
higher-spin massive particles by totally symmetric spinor of higher rank,
we derive the following equations for spin 1:
\begin{eqnarray}
&&\partial_\alpha F^{\alpha\mu} + {m\over 2} A^\mu = 0 \quad,\label{1a} \\
&&2 m F^{\mu\nu} = \partial^\mu A^\nu - \partial^\nu A^\mu \quad,
\label{2a}
\end{eqnarray}
In the meantime, in the textbooks, the latter set
is usually written as
\begin{eqnarray} &&\partial_\alpha F^{\alpha\mu} +
m^2 A^\mu = 0 \quad, \label{3a}\\ && F^{\mu\nu} = \partial^\mu A^\nu -
\partial^\nu A^\mu \quad, \label{4a}
\end{eqnarray}
The set
(\ref{3a},\ref{4a}) is obtained from (\ref{1a},\ref{2a}) after the
normalization change $A_\mu \rightarrow 2m A_\mu$ or $F_{\mu\nu}
\rightarrow {1\over 2m} F_{\mu\nu}$.  Of course, one can investigate other
sets of equations with different normalization of the $F_{\mu\nu}$ and
$A_\mu$ fields. Are all these sets of equations equivalent? -- I asked
in a recent series of my papers.

Ogievetski\u{\i} and Polubarinov argued that in the massless limit
``the system of $2s+1$ states is no longer irreducible; it decomposes
and describes a set of different particles with zero mass and
helicities $\pm s$, $\pm (s-1)$,\ldots $\pm 1$, $0$ (for integer spin and
if parity is conserved; the situation is analogous for half-integer
spins)." Thus, they did in fact contradict the Weinberg theorem.
But, in~\cite{DvoNo} I presented explicit forms of 4-vector potentials
and of parts of the antisymmetric tensor (AST) field and concluded that
the question should be solved on the basis of the analysis of
normalization issues. Here they are in the momentum representation:
\begin{equation}
u^\mu
({\bf p}, +1)= -{N\over \sqrt{2}m}\pmatrix{p_r\cr m+ {p_1 p_r \over
E_p+m}\cr im +{p_2 p_r \over E_p+m}\cr {p_3 p_r \over
E_p+m}\cr}\,,\quad  u^\mu ({\bf p}, -1)= {N\over
\sqrt{2}m}\pmatrix{p_l\cr m+ {p_1 p_l \over E_p+m}\cr -im +{p_2 p_l \over
E_p+m}\cr {p_3 p_l \over E_p+m}\cr}\,,\label{vp12}
\end{equation}
\begin{equation}
u^\mu ({\bf p}, 0) = {N\over m}\pmatrix{p_3\cr {p_1 p_3 \over E_p+m}\cr
{p_2 p_3 \over E_p+m}\cr m + {p_3^2 \over E_p+m}\cr}\,,\quad
u^\mu ({\bf p}, 0_t) = {N \over m} \pmatrix{E_p\cr p_1
\cr p_2\cr p_3\cr}\,\label{vp3}
\end{equation}
and
\begin{eqnarray}
{\bf B}^{(+)} ({\bf p},
+1) &=& -{iN\over 2\sqrt{2}m} \pmatrix{-ip_3 \cr p_3 \cr ip_r\cr} =
+ e^{-i\alpha_{-1}} {\bf B}^{(-)} ({\bf p}, -1 ) \quad,\quad   \label{bp}\\
{\bf B}^{(+)} ({\bf
p}, 0) &=& {iN \over 2m} \pmatrix{p_2 \cr -p_1 \cr 0\cr} =
- e^{-i\alpha_0} {\bf B}^{(-)} ({\bf p}, 0) \quad,\quad \label{bn}\\
{\bf B}^{(+)} ({\bf p}, -1)
&=& {iN \over 2\sqrt{2} m} \pmatrix{ip_3 \cr p_3 \cr -ip_l\cr} =
+ e^{-i\alpha_{+1}} {\bf B}^{(-)} ({\bf p}, +1)
\quad,\quad\label{bm}
\end{eqnarray}
and
\begin{eqnarray}
{\bf E}^{(+)} ({\bf p}, +1) &=&  -{iN\over 2\sqrt{2}m} \pmatrix{E_p- {p_1
p_r \over E_p +m}\cr iE_p -{p_2 p_r \over E_p+m}\cr -{p_3 p_r \over
E+m}\cr} = + e^{-i\alpha^\prime_{-1}}
{\bf E}^{(-)} ({\bf p}, -1) \quad,\quad\label{ep}\\
{\bf E}^{(+)} ({\bf p}, 0) &=&  {iN\over 2m} \pmatrix{- {p_1 p_3
\over E_p+m}\cr -{p_2 p_3 \over E_p+m}\cr E_p-{p_3^2 \over
E_p+m}\cr} = - e^{-i\alpha^\prime_0} {\bf E}^{(-)} ({\bf p}, 0)
\quad,\quad\label{en}\\
{\bf E}^{(+)} ({\bf p}, -1) &=&  {iN\over
2\sqrt{2}m} \pmatrix{E_p- {p_1 p_l \over E_p+m}\cr -iE_p -{p_2 p_l \over
E_p+m}\cr -{p_3 p_l \over E_p+m}\cr} = + e^{-i\alpha^\prime_{+1}} {\bf
E}^{(-)} ({\bf p}, +1) \quad,\quad\label{em}
\end{eqnarray}
where we denoted a normalization factor appearing in the
definitions of the potentials (and/or in the definitions of the physical
fields through potentials) as $N$ (which can, of course, be chosen
in arbitrary way, not necessarily as to be proportional
to $m$)\footnote{The possibility of appearance of additional
mass factors in commutation relations was also analyzed by us
in the recent series of papers.}
and $p_{r,l} = p_1 \pm ip_2$.  Thus, we find
that in the massless limit we may have in general divergent parts of
4-potentials and AST field, thus prohibiting to set $m=0$ in the equations
(\ref{1a}-\ref{4a}).  They are usually removed by ``electrodynamic" gauge
transformations. But, it was shown that the Lagrangian constructed from
the $(1,0)\oplus (0,1)$ (or antisymmetric tensor) fields admits another
kind of ``gauge" transformations, namely $F_{\mu\nu}  \rightarrow
F_{\mu\nu} +\partial_\nu \Lambda_\mu - \partial_\mu \Lambda_\nu$), with
some ``gauge" vector functions $\Lambda_\mu$. This becomes the origin of
the possibility of obtaining the quantum states (particles?) of different
helicities in both the $(1/2,1/2)$ and $(1,0)\oplus (0,1)$
representations.

In our formulation of generalized Maxwell equations these
in-general divergent terms can be taken into account explicitly,
thus giving additional terms in (\ref{1}-\ref{4}). As suggested
in~\cite{add} they may be applied to explanations
of several cosmological  puzzles. The detailed analysis of contradictions
between the Weinberg theorem and the
Ogievetski\u{\i}-Polubarinov-Kalb-Ramond conclusion (and also
discussions of~\cite{ml}) will be given in a separate publication.
Here, I would only like to mention few assumptions, under which Weinberg
derived his famous theorem:

\begin{itemize}

\item
The derivation is based on the analysis of the proper Lorentz
transformations only. The discrete symmetry operations of the full
Poincare Group (which, for instance, may lead to the change of
the sign of the energy) have not been considered there.

\item
The derivation assumes the particular choice of the coordinate frame,
namely $p_{1,2}=0$ and $p_3 =\vert {\bf p}\vert$.\footnote{As one can see,
unpolarized classical ${\bf E}$ and ${\bf B}$ depend indeed on the choice
of the coordinate system, Eq. (\ref{bn},\ref{en}).}

\item
The derivation does {\it not} assume that the antisymmetric tensor field
is related to 4-vector fields by certain derivative operator.

\end{itemize}

Finally, the intrinsic angular momentum operator of the electromagnetic
field (which can be found on the basis of the Noether theorem)
contain the coefficient functions which belong to different
representations of the Lorentz Group, $\vec{\bf S} \sim \vec{\bf E}\times
\vec{\bf A}$ and it acts in the Fock space~\cite{DvoNo}.  Furthermore, the
condition (35) $W^\mu =kp^\mu$ is {\it not} the only condition which can
be imposed for massless particles. Namely, as stressed in~[15c] the
Pauli-Lubanski vector {\it may} be a space-like vector in this case, what
would correspond to ``infinite spin" representation.

Finally, we would like to add some words to the Dirac derivation of
equations (30-33) of the Gersten paper and their analysis. We derived the
formula (for spin 1)
\begin{equation}
\left [ \vec{\bf S}^i  (\vec {\bf S}\cdot
\vec{\bf p}) \right ]^{jm} = \left [ \vec{\bf p}^i I^{jm} - i [\vec{\bf
S}\times \vec{\bf p}]^{i,jm} - p^m \delta^{ij} \right ]\,,
\end{equation}
with $i$ being the vector index and $j,m$ being the matrix indices.
Hence, from the equation ($k=1$)
\begin{equation}
\left\{ kp_{t}+S_{x}p_{x}+S_{y}p_{y}+S_{z}p_{z}\right\}
{\bbox \psi} =0,  \label{b1}
\end{equation}
multiplying subsequently by $S_x$, $S_y$ and $S_z$ one can obtain
in the case $S=1$
\begin{equation}
\left\{ p_{x}+S_{x}p_{t}-iS_{y}p_{z}+iS_{z}p_{y}\right\}^{jm}
{\bbox\psi}^m
- (\vec{\bf p}\cdot \vec{\bbox\psi}) \delta^{xj}=0,
\label{a2}
\end{equation}
\begin{equation}
\left\{ kp_{y}+S_{y}p_{t}-iS_{z}p_{x}+iS_{x}p_{z}\right\}^{jm}
{\bbox\psi}^m - (\vec{\bf p}\cdot \vec{\bbox\psi}) \delta^{yj}=0,
\label{a3} \end{equation}
\begin{equation}
\left\{ kp_{z}+S_{z}p_{t}-iS_{x}p_{y}+iS_{y}p_{x}\right\}^{jm}
{\bbox\psi}^m
- (\vec{\bf p}\cdot \vec{\bbox\psi}) \delta^{zj}=0.
\label{a4}
\end{equation}
One can see from the above that the equations (31-33) of ref.~\cite{Gers}
can be considered as the consequence of the equation (30) and additional
``transversality condition" $\vec{\bf p}\cdot \vec{\bbox\psi} =0$ in the
case of the spin-1 consideration.  So, it is not surprising that they are
equivalent to the complete set of Maxwell's equations.
They are obtained  after multiplications by corresponding ${\bf S}$
matrices.  But, the crucial mathematical problem with such a
multiplication is that the ${\bf S}$ matrices for boson spins are {\it
singular}, $\mbox{det} S_{x} =\mbox{det} S_y =\mbox{det} S_z \equiv 0$,
which makes the above procedure doubtful\footnote{After
the analysis of the literature I learnt that, unfortunately,
a similar procedure has been applied in the derivation of many
higher-spin equations without proper explanation and precaution.}
and leaves room for possible generalizations. Moreover,
the right hand side of the equation (30) of~[1] may also be different from
zero according to our analysis above.

The conclusion of my paper is: unfortunately,
the possible consequences following
from Gersten's equation (9) have not been explored in full; on this basis
we would like to correct his conclusion and his claim in the abstract
of~[1].  -- It is the {\it generalized} Maxwell equations (many versions
of which have been proposed during the last 100 years, see, for
instance,~\cite{DvoR}) that should be used as a guideline for proper
interpretations of quantum theories.

\acknowledgments
I greatly appreciate discussions with
Profs. A. Chubykalo, L. Horwitz and A. Gersten, and the useful
information  from Profs. D. Ahluwalia, A. F. Pashkov, E. Recami
and M. Sachs.

Zacatecas University, M\'exico, is thanked for awarding
the professorship.  This work has been partly supported by
the Mexican Sistema Nacional de Investigadores and the Programa
de Apoyo a la Carrera Docente.

\end{document}